\documentclass[10pt,aps,prd,amsmath,amssymb,superscriptaddress,twocolumn,nofootinbib,showpacs,preprintnumbers,amsfonts, notitlepage]{revtex4-2}

\usepackage{hyperref}
\usepackage{amsmath,amssymb}
\usepackage{lmodern,bm}
\usepackage{multirow}
\usepackage{xspace}
\usepackage{tabularx}
\usepackage{colortbl}
\usepackage{tikz}
\usepackage{color}
\usepackage{comment}
\usepackage{standalone}

\usepackage{pifont}

\DeclareUnicodeCharacter{2212}{-}
\DeclareUnicodeCharacter{2032}{'}
\DeclareUnicodeCharacter{2261}{\equiv}
\DeclareUnicodeCharacter{2217}{*}

\usepackage[compat=1.1.0]{tikz-feynman}

\providecommand{\U}[1]{\protect\rule{.1in}{.1in}}
\newcommand{\mevnospace}{\ensuremath{{\mathrm{Me\kern -0.1em V}}}}
\newcommand{\gev}{\ensuremath{{\mathrm{\,Ge\kern -0.1em V}}}\xspace}
\newcommand{\mev}{\ensuremath{{\mathrm{\,Me\kern -0.1em V}}}\xspace}
\newcommand{\kev}{\ensuremath{{\mathrm{\,ke\kern -0.1em V}}}\xspace}
\newcommand{\fm}{\ensuremath{{\mathrm{\,fm}}}\xspace}

\newcommand{\pythia}{{\tt Pythia~8}\xspace}

\usepackage{mfirstuc} 
\newcommand{\addReviewer}[2]{
  \expandafter\newcommand\csname #1\endcsname[1]{{\textbf{ \color{#2} \capitalisewords{#1}:\,##1}}}
  \expandafter\newcommand\csname #1cor\endcsname[2]{{\color{#2} \capitalisewords{#1}:\,\st{##1}{\textbf{##2}}}}
  \expandafter\newcommand\csname #1color\endcsname{#2}
  \expandafter\newcommand\csname #1todo\endcsname[1]{{\todo[inline,color=white!70!#2, caption={}]{\textbf{\capitalisewords{#1}}: ##1}}}
}

\usepackage{soul,color}
\definecolor{chromeyellow}{rgb}{1.0, 0.65, 0.0}

\addReviewer{ale}{chromeyellow}
\addReviewer{ut}{red}
\addReviewer{dm}{green}

\newcommand{\be}{\begin{equation}}
\newcommand{\ee}{\end{equation}}

\begin{document}

\title{Production of loosely-bound hadron molecules from bottomonium decays}

\newcommand{\unito}{Dipartimento di Fisica, Universit\`a degli Studi di Torino, Via Pietro Giuria, 1, I-10125 Torino, Italy}
\newcommand{\infnto}{INFN Sezione di Torino, Via Pietro Giuria 1, I-10125 Torino, Italy}
\newcommand{\mift}{Dipartimento di Scienze Matematiche e Informatiche, Scienze Fisiche e Scienze della Terra,
Universit\`a degli Studi di Messina, Viale Ferdinando Stagno d'Alcontres 31, I-98166 Messina, Italy}
\newcommand{\catania}{INFN Sezione di Catania, Via Santa Sofia 64, I-95123 Catania, Italy}

\author{Davide Marietti}
\affiliation{\unito}
\author{Alessandro Pilloni}
\affiliation{\mift}
\affiliation{\catania}
\author{Umberto Tamponi}
\affiliation{\infnto}

\begin{abstract}
We present multiple results on the production of loosely-bound molecules in bottomonium annihilations and $e^+e^-$ collisions at $\sqrt{s} = 10.58\gev$.
We perform the first comprehensive test of several models for deuteron production against all the existing data in this energy region. We fit the free parameters of the models to reproduce the observed cross sections, and we predict the deuteron spectrum and production and the cross section for the  $e^+e^- \to d\bar{d} + X$ process both at the $\Upsilon(1,2,3S)$ resonances and at $\sqrt{s}=10.58\gev$. The predicted spectra show differences but are all compatible with the uncertainties of the existing data. These differences could be addressed if larger datasets are collected by the Belle~II experiment.
Fixing the source size parameter to reproduce the  deuteron data, we then predict the production rates for $H$ dibaryon and hypertriton in this energy region using a simple coalescence model. Our prediction on $H$ dibaryon production rate is below the limits set by the direct search at the Belle experiment, but in the range accessible to the Belle~II experiment.
The systematic effect due to the MC modelling of quarks and gluon fragmentation into baryons is reduced deriving a new tuning of the \pythia MonteCarlo generator using the available measurement of single- and double-particle spectra in $\Upsilon$ decays.
\end{abstract}

\maketitle

\section{Introduction}
\label{sec:intro}

Deuteron is the simplest example of a hadron molecule, {\em i.e.} a composite state of two or more hadrons bound by color-neutral (residual) strong interactions. The phenomenology related to its production in high energy processes can shed light to several highly-debated problems in contemporary particle physics. 
Its binding energy $B_d = 2.22\mev$ is much smaller than the mass of the pion that mediates the interaction between the constituents; this entails for some universal properties to hold~\cite{Weinberg:1962hj,Guo:2017jvc}. In the field of exotic spectroscopy, deuteron can be used as a well understood reference to predict properties of other hadronic molecule candidates. A better understanding of its formation mechanism would allow for precise, falsifiable predictions on the production rate of heavy hadron molecules which can then be tested against existing data. 
Antideuteron ($\bar{d}$) plays an important role in astroparticle physics as well: its presence in cosmic rays has been proposed as a low background detection channel for dark matter indirect searches~\cite{Donato:1999gy}, initiating vast and ongoing theoretical~\cite{Aramaki:2015pii,Korsmeier:2017xzj,Serksnyte:2022onw} and experimental~\cite{Battiston:2008zza,vonDoetinchem:2020vbj,Abe:2017yrg,Aramaki:2015laa} efforts. Reducing the uncertainties on its production rate in high energy collisions is crucial to improve the modeling of both the standard model backgrounds 
and the dark matter-induced signal. 

Looking at the experimental side, $\bar{d}$ production has been observed at colliders since the early 1960's~\cite{Hagedorn:1960zz, Butler:1963pp}. We have by now collected measurements in several different processes: 
$p\overset{\scriptscriptstyle(-)}{p}$~\cite{ALICE:2015wav, ALICE:2017xrp,ALICE:2019dgz,ALICE:2020foi,ALICE:2021mfm,Alexopoulos:2000jk,British-Scandinavian-MIT:1977tan}, proton-nucleus~\cite{Schwarzschild:1963zz,ALICE:2019bnp}, nucleus-nucleus collisions~\cite{ALICE:2015wav,ALICE:2017nuf,PHENIX:2004vqi,STAR:2001pbk,Bearden:2000we,E864:2000loc,E864:2000auv,NA52NEWMASS:1996ptr,NA44:1999hqx}, $\Upsilon(nS)$~\footnote{Hereafter we collectively call $\Upsilon(nS)$ the three narrow $\Upsilon$ resonances below the $B\bar B$ threshold: $\Upsilon(1S)$, $\Upsilon(2S)$, and $\Upsilon(3S)$.} annihilations~\cite{ARGUS:1985cfz,ARGUS:1989sto,CLEO:2006zjy,BaBar:2014ssg}, $e^+e^-$ scattering at $\sqrt{s} = 10.580\gev$ and at the $Z^0$ pole~\cite{BaBar:2014ssg,ALEPH:2006qoi}, photoproduction~\cite{H1:2004maq}, and deep inelastic scattering~\cite{ZEUS:2007bxq}.
Explaining all these reactions quantitatively is  challenging because of the wide range of energies involved, from a binding energy of few \mevnospace's to production energy scales of $\sim O(1\text{--}100\gev)$ depending of the process. The problem has been studied in a number of phenomenological models~\cite{Esposito:2015fsa, Esposito:2020ywk} that relate the production rate and kinematics of a hadron molecule to that of its constituents. Modelling the correlations between the constituents is thus of primary importance.

In this work we will focus on the measurements performed by $e^+e^-$ colliders in the bottomonium region. As mentioned, experimental results are available at the three narrow $\Upsilon(nS)$ peaks, and in $e^+e^-$ collisions off-peak. The availability of measurements both on- and off-peak allows us to study the depencence on the parton process and indirectly the production of antideuteron: $\Upsilon(nS)$ annihilate mostly into gluons, while outside the resonance the quark-antiquark production via  $e^+e^- \to u\bar{u}, d\bar{d}, s\bar{s}, c\bar{c}$ (collectively referred as the $e^+e^- \to q \bar{q}$ continuum) dominates the hadronic cross section. 
Despite the abundance of data, little attention has been paid to this sector so far. The bulk of theoretical activity has been focused on describing the high statistics data in $pp$ and PbPb collisions. The bottomonium sector shows some peculiar characteristics that makes it particularly interesting in our opinion. $\Upsilon(nS)$ annihilations are approximately seven times more likely to produce $\bar{d}$ than the $e^+e^- \to q\bar{q}$ continuum at similar energies, pointing to strong differences between gluon and quark fragmentation into baryons, which is usually not addressed in the theoretical works describing antideuteron formation. Moreover, the size of the interaction region (smaller than in ion collisions), together with the limited track multiplicity per event, reduce the complications due to multiple rescattering~\cite{Esposito:2020ywk}. 

The first part of this paper is devoted to testing several coalescence-inspired models discussed in the literature, applying them for the first time to all the recent data coming from $e^+e^-$ colliders in the bottomonium region. We fit the parameters of each model to the observed cross sections, and benchmark them against the $\bar{d}$ spectra and the $d\bar{d}$ associated production rate. 
In the second part we use a coalescence model tuned on antideuteron to predict the production rate of other loosely-bound hadron molecules: the hypertriton and a shallow $H$ dibaryon.



\section{Models and experimental measurements}
\label{sec:models}

Antinuclei production in high energy collision is usually described by two classes of models:  thermal~\cite{Andronic:2017pug, Bellini:2018epz} and coalescence ones.  We focus on the latter, in which the $\bar{d}$ production is described as a binding process between nucleons that happen to be created close to each other in both coordinate and momentum space. In $e^+e^-$ collisions the spatial distance of the two nucleons is expected to be negligible,\footnote{The source radius of $e^+e^-$ collisions has never been measured at $\sqrt{s}\approx 10\gev$. However, measurement of Bose-Einstein correlations at LEP indicate an effective radius of $\simeq 0.7\fm$ at $\sqrt{s}\approx 200\gev$~\cite{L3:2002ess}.} and the binding probability $P(\bar{p}\bar{n} \to \bar{d}X)$ is only function of their relative momentum in the center-of-mass (CoM) frame $k = \frac{1}{2}|\bar{p}_p - \bar{p}_n|_\text{CoM}$~\cite{Butler:1963pp, Schwarzschild:1963zz}.

We consider three different models for the binding probability. In model A,  the $\bar{d}$ formation probability is a step function~\cite{Gustafson:1993mm, Artoisenet:2010uu},
\begin{equation}\label{eq:prob_simple}
    P(\bar{p} \bar{n} \rightarrow \bar{d} X \, | \, k) =
    \begin{cases}
        1 \hspace{1cm} k \leq k_\text{cut}\\
        0 \hspace{1cm} k > k_\text{cut}
    \end{cases},
\end{equation}
where $k_\text{cut}$ is the theory parameter, function of the color string breakup length $\sigma_s$ and of the binding energy $B_d$:
\begin{equation}\label{eq:sigma}
        (2k_\text{cut})^3 =  \frac{36}{\sqrt{\pi}} \sigma_s^{-2} \sqrt{m_{p} B_{d}}\, .
\end{equation}
Since we cannot estimate $\sigma_s$ independently, we will treat $k_\text{cut}$ as a free parameter, fitted to data to reproduce the observed cross sections. Typical values reported by other authors range from $70$ to $200\mev$ according to the process~\cite{Ibarra:2012cc}.

Model B introduces a microscopic picture based on the Wigner function representation~\cite{Kachelriess:2019taq}. The binding probability is given by:
\begin{equation}\label{eq:prob_two_g}
    P(\bar{p} \bar{n} \rightarrow \bar{d} X \, | \, k) = 3 \Big{(} \zeta_1 (\sigma) \Delta  e^{-k^2 d_1^2} + \zeta_2 (\sigma) (1-\Delta) e^{-k^2 d_2^2} \Big{)},
\end{equation}
where $\zeta_i(\sigma)$ is a known function of the source size $\sigma$, $d_i$ are coefficients describing the $\bar{d}$ wave function, and $\Delta=0.581$. Again, we consider $\sigma$ as a fit parameter.\footnote{In principle $\sigma$ could be independently constrained by studying two-particle correlations, thus fixing all the parameters of the model. Such studies unfortunately have not been performed yet.}

Model C describes the $\bar{d}$ formation as a dynamical process with probability given by the incoherent sum of the cross sections of known exclusive processes producing $\bar{d}$, $\sigma[\bar{N_1}\bar{N_2} \rightarrow \bar{d}  X](k)$, with $\bar{N} = \{\bar{p}, \bar{n}\}$ and $X = \{\gamma, \pi^0, ...\}$. The sum of these cross sections is then normalized by a free parameter $\sigma_0$~\cite{Dal:2015sha}, leading to
\begin{equation}\label{eq:prop_sez}
    P(\bar{N_1}\bar{N_2} \rightarrow \bar{d} X \, | \, k) = \frac{\sum \sigma[\bar{N_1}\bar{N_2} \rightarrow \bar{d} X](k)}{\sigma_0}.
\end{equation}
All these models have been applied to at least one of the measurements from $e^+e^-$ colliders by the original authors, but no global fit has been attempted so far.

As mentioned before, antideuteron production in the bottomonium region was first observed by Argus~\cite{ARGUS:1985cfz,ARGUS:1989sto} and then in further studies by  CLEO~\cite{CLEO:2006zjy} and BaBar~\cite{BaBar:2014ssg}. We do not consider the Argus measurements, since they are performed summing over all the  $\Upsilon(nS)$ datasets, have sizable statistical  uncertainties and appear systematically shifted from most recent and precise measurements. CLEO reported the first observation of $\Upsilon(1S) \to \bar{d}  X$, a low-statistics measurement of $\Upsilon(2S) \to \bar{d}  X$, an upper limit for $e^+e^- \to \bar{d}  X$ continuum production, and three candidate events where both $d$ and $\bar{d}$ are produced. BaBar measured $\mathcal{B}\left(\Upsilon(nS) \to \bar{d}  X\right)$ for all narrow bottomonia and observed for the first time production from the continuum. The latter is measured at the $\Upsilon(4S)$ peak, assuming the contribution from $B^{0/\pm} \to \bar{d}  X$ to be negligible.~\footnote{This seems reasonable, given the upper limit ${\cal B}\left(B^{0} \to pp \bar{p} \bar{p}\right) < 2 \times 10^{-7}$~\cite{BaBar:2018gjv}.} No search for associated production is reported. All the experiments published the $\bar{d}$ momentum spectrum.
The measurements used in our analysis are summarized in Table~\ref{tab:exp_BR}. 
\renewcommand{\arraystretch}{1.7}
\setlength{\tabcolsep}{4pt}
\begin{table}[h!]
\begin{center}
\begin{tabular}{ccc}
    \hline
    Process & CLEO $(\times 10^{-5})$  & BaBar $(\times 10^{-5})$ \\
    \hline\hline

    $\mathcal{B}$($\Upsilon(1S)\rightarrow \bar{d}X$) & \small{$2.86 \pm 0.19 \pm 0.21$} & \small{$2.81 \pm 0.49^{+0.20}_{-0.24}$} \\
    $\mathcal{B}$($\Upsilon(2S)\rightarrow \bar{d}X$) & \small{$3.37 \pm 0.50 \pm 0.25$} &  \small{$2.64 \pm 0.11^{+0.26}_{-0.21}$} \\ 
    $\mathcal{B}$($\Upsilon(3S)\rightarrow \bar{d}X$) & --- & \small{$2.33 \pm 0.15^{+0.31}_{-0.28}$} \\
    $\frac{\sigma(e^+e^-\rightarrow \bar{d}X)}{\sigma(e^+e^-\rightarrow {\rm hadrons})}$ & \small{$< 1$}  & \small{$0.301 \pm 0.013^{+0.037}_{-0.031} $} \\[0.3ex]
    \hline
\end{tabular}\end{center}
\caption{Available data about $\bar{d}$ inclusive production in the bottomonium region~\cite{BaBar:2014ssg, CLEO:2006zjy}. The first uncertainty is statistical, the second is systematic.}
\label{tab:exp_BR}
\end{table}

\section{Tuning of the MonteCarlo generators}
\label{sec:MC}
As showed in the previous section, the models provide a coalescence probability depending on $k$. This has to be multiplied by the two-nucleon cross section $\frac{d \sigma}{dk}$, representing the probability of two nucleons being produced with relative momentum $k$, that encodes the dynamical correlations between nucleons arising from the hadronization process. Unfortunately, no measurement of this cross sections is available. For this reason we rely on MonteCarlo (MC) simulation libraries, in our case \pythia~\cite{Sjostrand:2014zea}, to model it. The results obtained in this way obviously depend on the choice of the generator and its specific tuning~\cite{Dal:2012my, Dal:2014nda}. For this reason, we first develop a \pythia tuning set optimized for the description of $e^+e^-$ scattering in the bottomonium region.
The simulation is performed using the Belle~II Analysis Software Framework ({\tt basf2})~\cite{basf2,basfonline}, which offers a convenient interface to several generators. We use {\tt Evtgen}~\cite{evtgen} to simulate the  $\Upsilon(nS)$ decays, and {\tt KK}~\cite{Jadach:2000ir} for the continuum. The fragmentation is then performed by \pythia for both.  
We develop two tunings, one for the $\Upsilon(nS)$ and one for the continuum, separately.
We start from the standard set used by the Belle~II collaboration, and we test it against a measurement of the the single-proton differential cross section $\sigma(e^+e^- \rightarrow p X)$ at $\sqrt{s} = 10.520\gev$~\cite{Belle:2015hut,Belle:2020pvy} and the total cross sections for hyperons and charmed baryons production at the same energy~\cite{visible_total_sez}.
\begin{figure}[b]
	\centering
    \includegraphics[width=\linewidth]{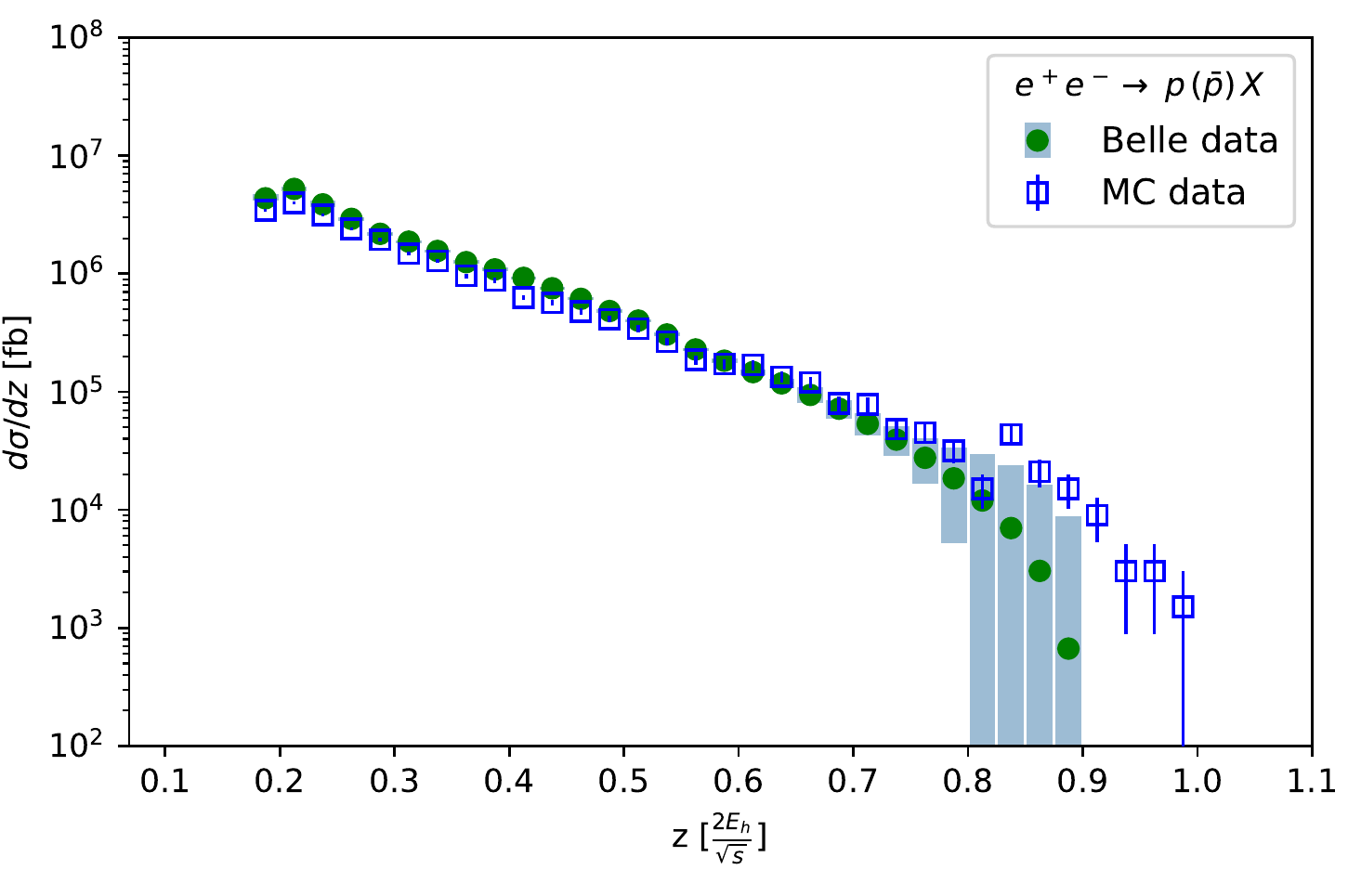}
    \caption{The $e^+e^- \rightarrow \overset{\scriptscriptstyle(-)}{p} X$ differential cross section (combining $p$ and $\bar{p}$ contributions), as a function of $z = 2 E_h / \sqrt{s}$. MC data (blue squares) are generated using \pythia with Belle~II tuning set. Notice the good agreement with experimental data (green points). Boxes and bars represent the systematic  and statistical uncertainties, respectively~\cite{Belle:2015hut,Belle:2020pvy}.}
    \label{fig:diff_cross_sez}
\end{figure}
\begin{figure*}[t]
	\centering
    \includegraphics[width=0.495\linewidth]{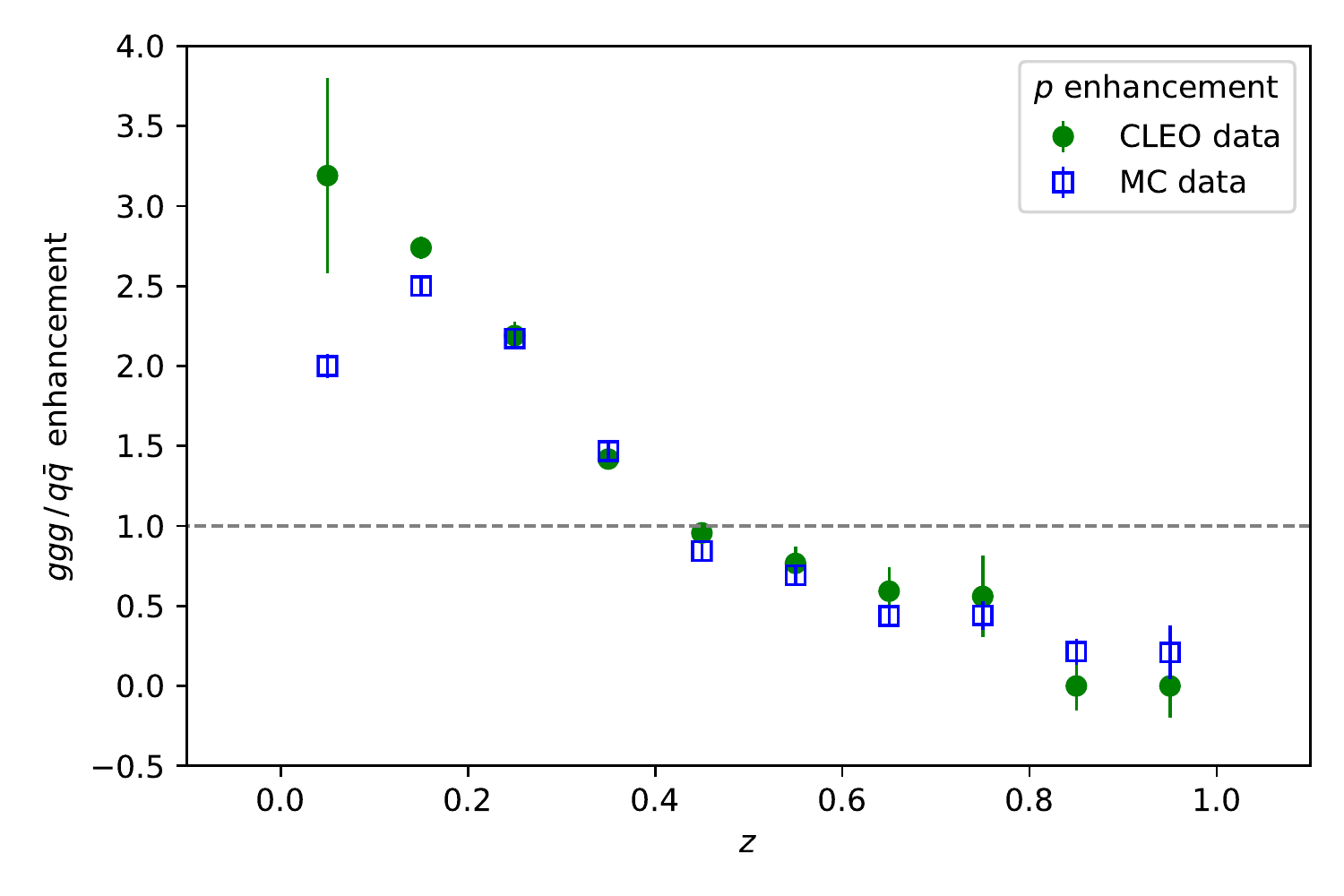}
    \includegraphics[width=0.495\linewidth]{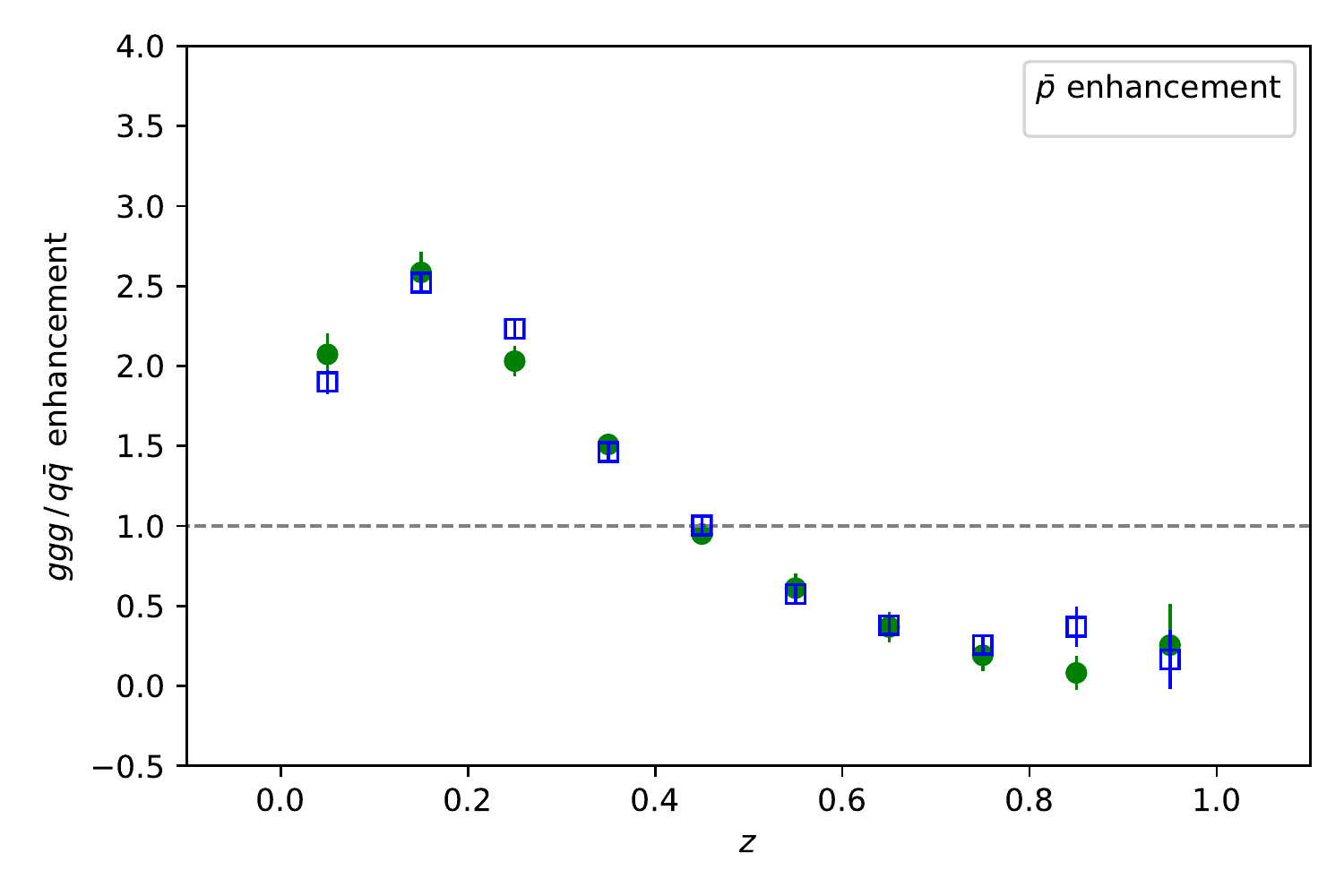}
    \includegraphics[width=0.495\linewidth]{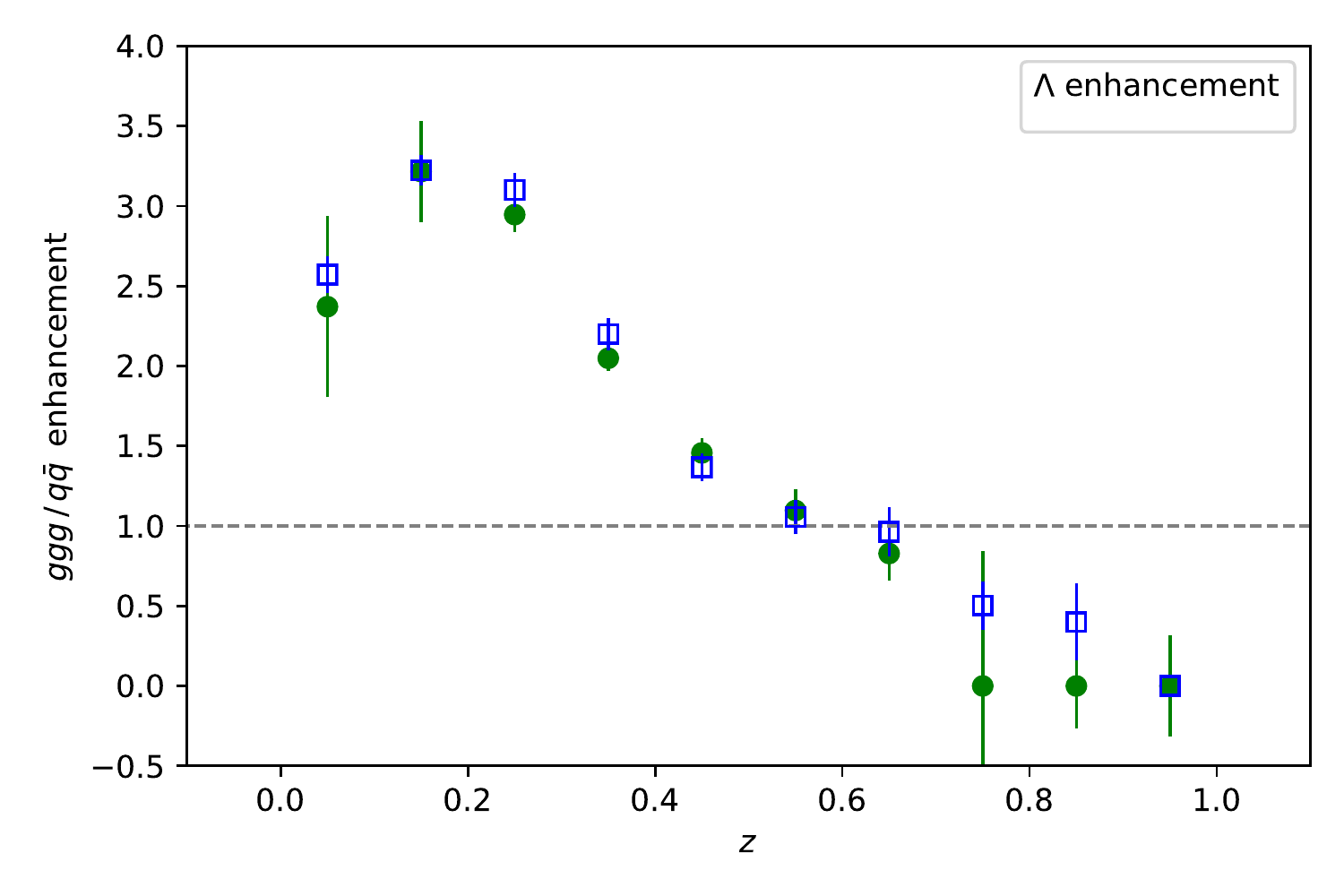}
    \includegraphics[width=0.495\linewidth]{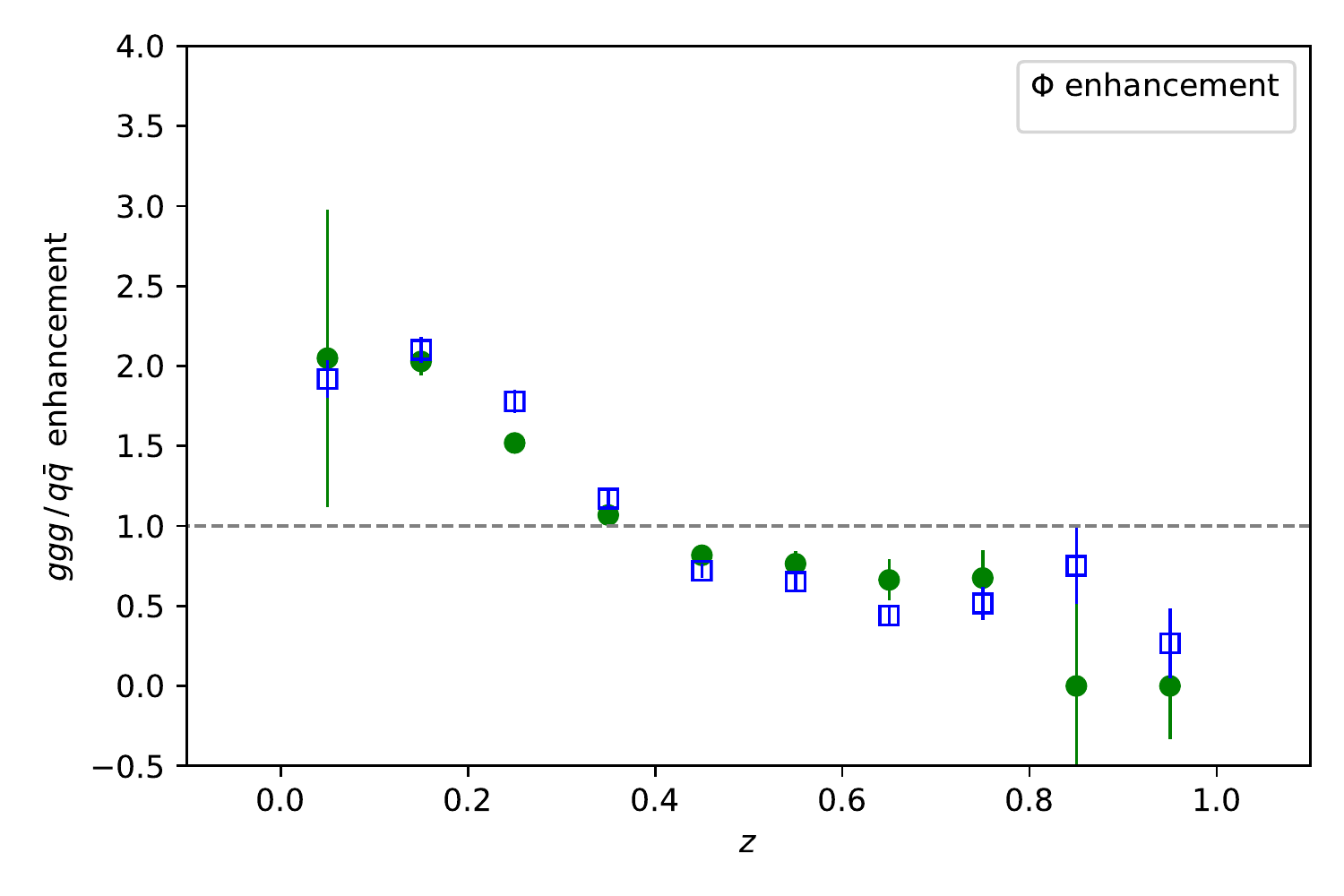}
    \caption{The enhancement as a function of momentum scaled $z$ for $\Lambda$, $p$, $\bar{p}$, $\phi$. The tuned MC (blue squares) is compared with the CLEO experimental data (green points)~\cite{enhance}.}
    \label{fig:enhance_scaled_tuned}
\end{figure*}
We find a reasonable agreement between the data and the simulation as shown in  Figure~\ref{fig:diff_cross_sez}, and we do not perform any further tuning to describe the continuum. 
For the $\Upsilon(nS)$ we compare with the measurement of the $ggg/q\bar{q}$ enhancement performed by CLEO~\cite{enhance}. We find disagreement, particularly at low momenta. 
\begin{table}[t]
\renewcommand{\arraystretch}{1.5}
\begin{tabular}{|l|}
    \hline
    {\tt StringZ:aLund = 0.22}\\
    {\tt StringZ:bLund = 1.35}\\
    {\tt StringZ:aExtraDiquark = 1.05}\\
    {\tt StringPT:sigma = 0.238}\\
    {\tt StringFlav:probQQtoQ = 0.091}\\
    {\tt StringFlav:probStoUD = 0.32}\\
    {\tt StringFlav:probSQtoQQ = 1.0}\\
    \hline
\end{tabular}
\caption{Values of the \pythia parameters tuned for the $\Upsilon(nS)$ decays.}
\label{tab:tuning}
\end{table}
We improve the simulation doing a grid tuning of the \pythia parameters related to the momentum and multiplicity of produced hadrons, the suppression of diquark over quark production, and the strangeness production. The optimized parameters are reported in Table~\ref{tab:tuning}. Figure~\ref{fig:enhance_scaled_tuned} shows the comparison between data and simulation after the tuning. As said, we use this setting for $\Upsilon(nS)$ fragmentation only.

\section{Study of the $\bar{d}$ production}
\label{sec:d_prod}
We produce $5 \times 10^7$ events for each $\Upsilon(1S)$, $\Upsilon(2S)$ and $\Upsilon(3S)$, and $10^8$ continuum events at $\sqrt{s} = 10.58$\gev using the settings described in the \hyperref[sec:MC]{previous section}.
Antinucleons from weak decays are produced outside of the source volume and we therefore  discard them, while the ones produced in the decay of strong resonances are kept~\cite{Gustafson:1993mm}.
Removing nucleons from hyperon decays roughly halves the predicted $\bar{d}$ production rate in $\Upsilon$ annihilations, while charmed baryons give a negligible contribution.
For each event containing a $\bar{p}\bar{n}$ pair, we decide whether a $\bar d$ is produced or not according to the probabilities discussed in Section~\ref{sec:models}. The procedure is repeated by varying the free parameter of each model. The yields are then compared with the experimental values obtaining a $\chi^2$ scan as a function of the model parameter. The position of the $\chi^2$ minimum and the  $\Delta\chi^2 = 1$ interval are chosen as best fit value and related uncertainty~\cite{statistics}.
Results are reported in Table~\ref{tab:best_fit}. 
\begin{table}[b]
\begin{center}
\begin{tabular}{lccc}
	\hline
	Process & $k_{cut}$ $[\mevnospace]$ & $\sigma$ [fm] & $1/\sigma_0$ [b$^{-1}$]\\
	\hline\hline
	$\Upsilon(1S)\rightarrow \bar{d} X$  & $75.1^{+2.4}_{-2.3}$ & $1.58^{+0.06}_{-0.05}$  & $1.87^{+0.22}_{-0.17}$\\
	$\Upsilon(2S)\rightarrow \bar{d} X$  & $75.5^{+2.3}_{-2.5}$ & $1.59^{+0.05}_{-0.05}$  & $1.84^{+0.16}_{-0.15}$\\
	$\Upsilon(3S)\rightarrow \bar{d} X$  & $71.6^{+2.0}_{-2.9}$ & $1.68^{+0.09}_{-0.08}$  & $1.57^{+0.16}_{-0.21}$\\
	$\Upsilon(nS)\rightarrow \bar{d} X$  & $73.7^{+1.3}_{-1.4}$ & $1.60^{+0.03}_{-0.02}$  & $1.75^{+0.13}_{-0.12}$\\
	$e^+e^-\rightarrow \bar{d} X$  & $63.8^{+3.2}_{-3.0}$ & $1.89^{+0.10}_{-0.09}$  & $1.13^{+0.14}_{-0.16}$\\
	\hline
\end{tabular}
\end{center}
\caption{The best-fit values of the phenomenological parameters $k_{cut}$ (model A), $\sigma$ (model B), and $1/\sigma_0$ (model C). The $\Upsilon(nS) \rightarrow \bar{d} X$ result is obtained with a simultaneous fit of all the $\Upsilon(nS)$ measurements.}
\label{tab:best_fit}
\end{table}

All the three models show consistent values for $\Upsilon(1S)$, $\Upsilon(2S)$ and $\Upsilon(3S)$  decays, and significantly different ones for the continuum. By itself, this is not enough to explain the difference between the rate of $\bar{d}$ observed in the two processes, pointing to significant differences in $ggg$ and $q\bar{q}$ fragmentation.
We also observe a possible dependence of the coalescence parameters from the center of mass energy for all the models. This effect is not statistically significant and might depend on the different fraction of $ggg$ events at the three $\Upsilon$ resonances, due to the presence of radiative transitions to $\chi_{bJ}(nP)$ that are not subtracted from the $\Upsilon(nS)$ decays.
In several cases we observe disagreement between our results and previous partial fits of the same datasets by other authors~\cite{Gustafson:1993mm, Ibarra:2012cc, Dal:2015sha, Artoisenet:2010uu}. We attribute this to the different generator choices and tunings.

\begin{figure*}[!t]
	\centering
    \includegraphics[width=0.495\textwidth]{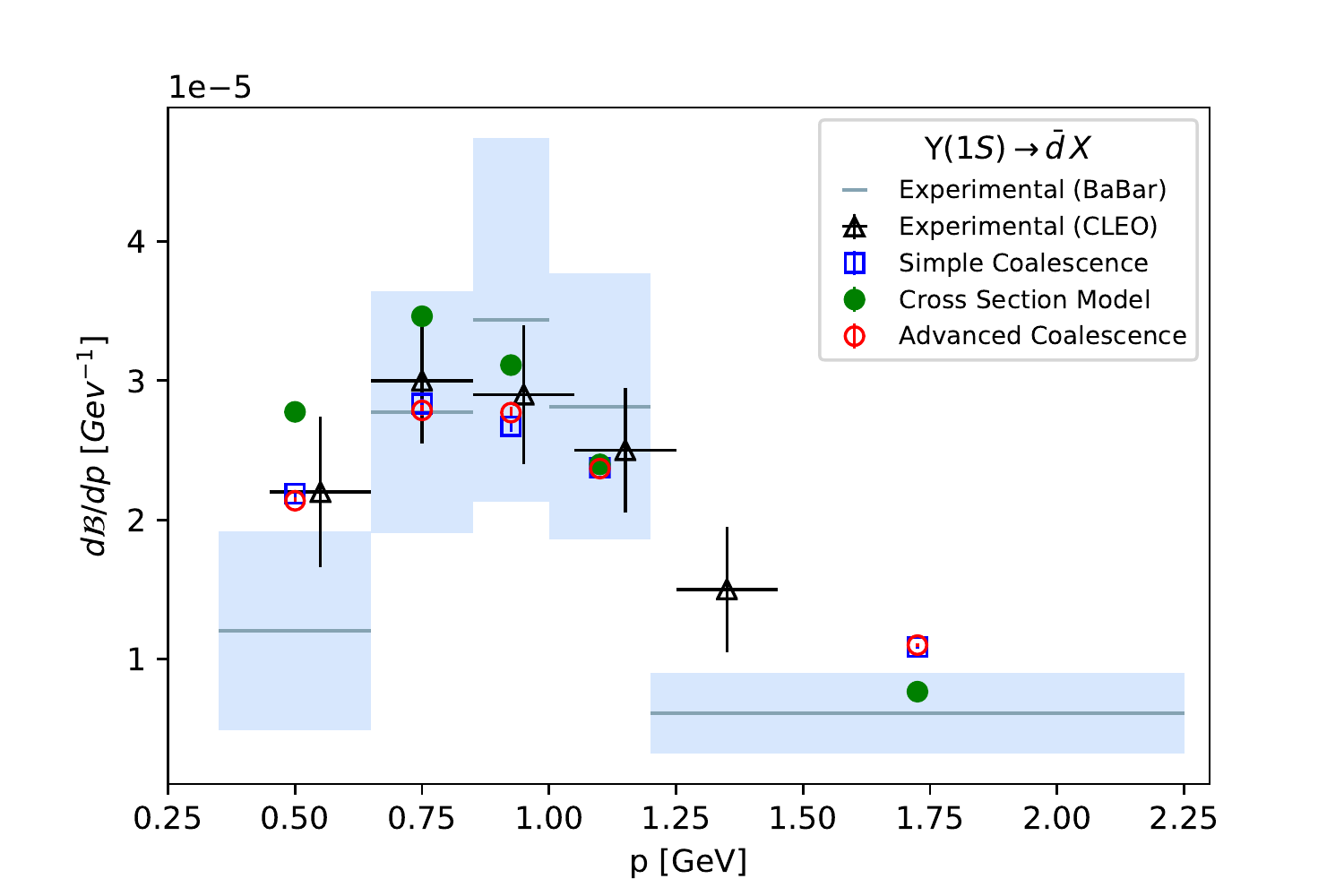}
    \includegraphics[width=0.495\linewidth]{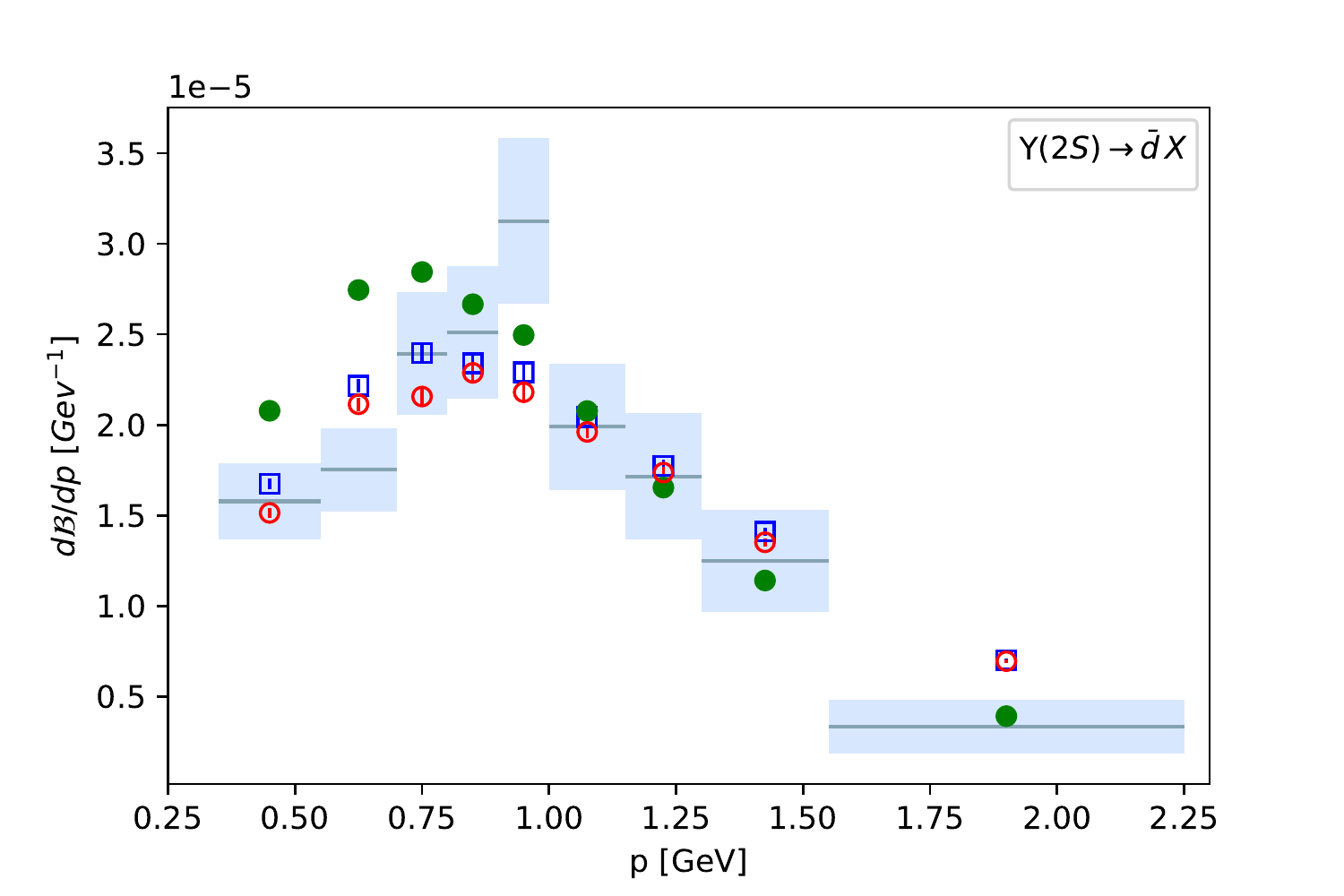}
    \includegraphics[width=0.495\linewidth]{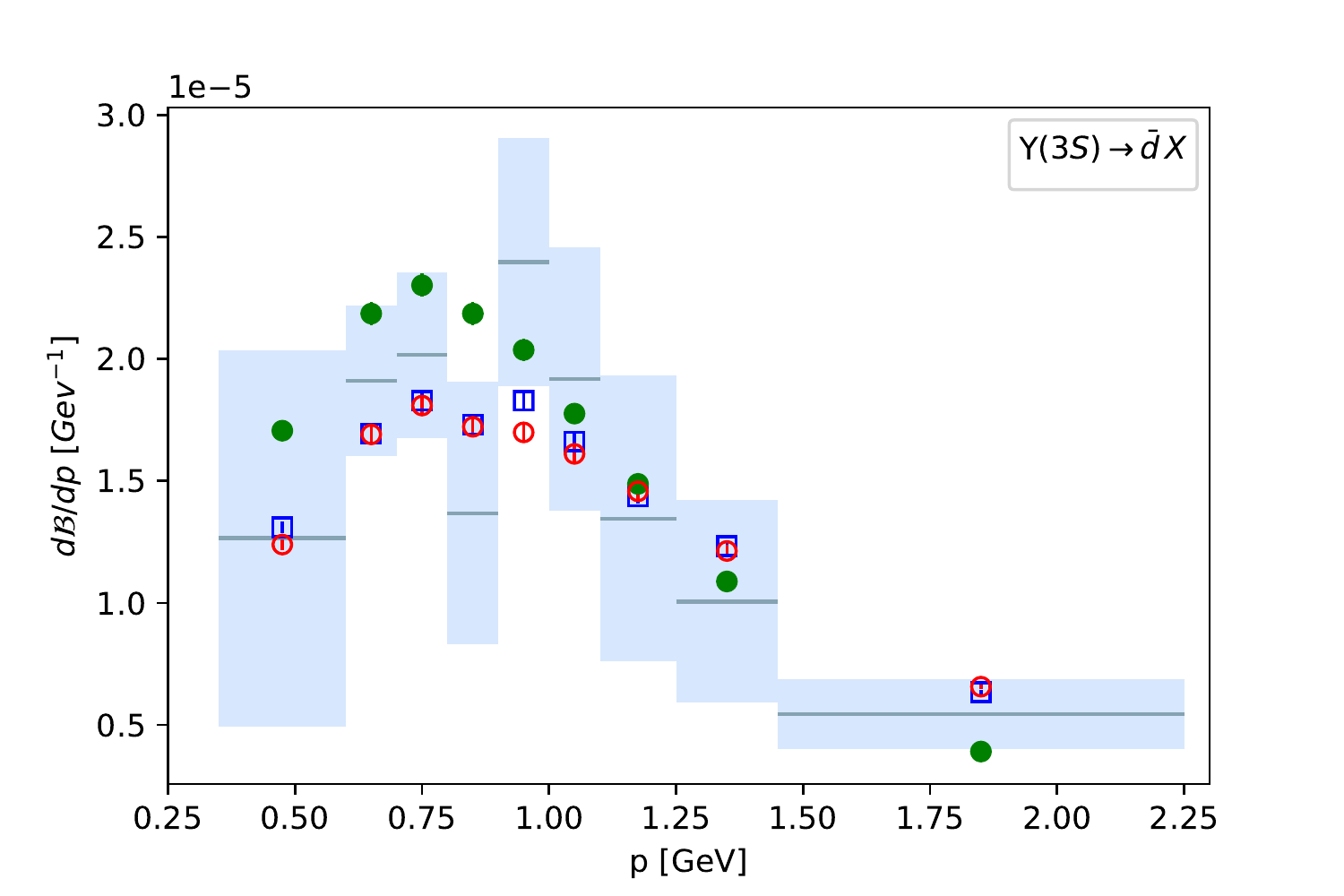}
    \includegraphics[width=0.495\linewidth]{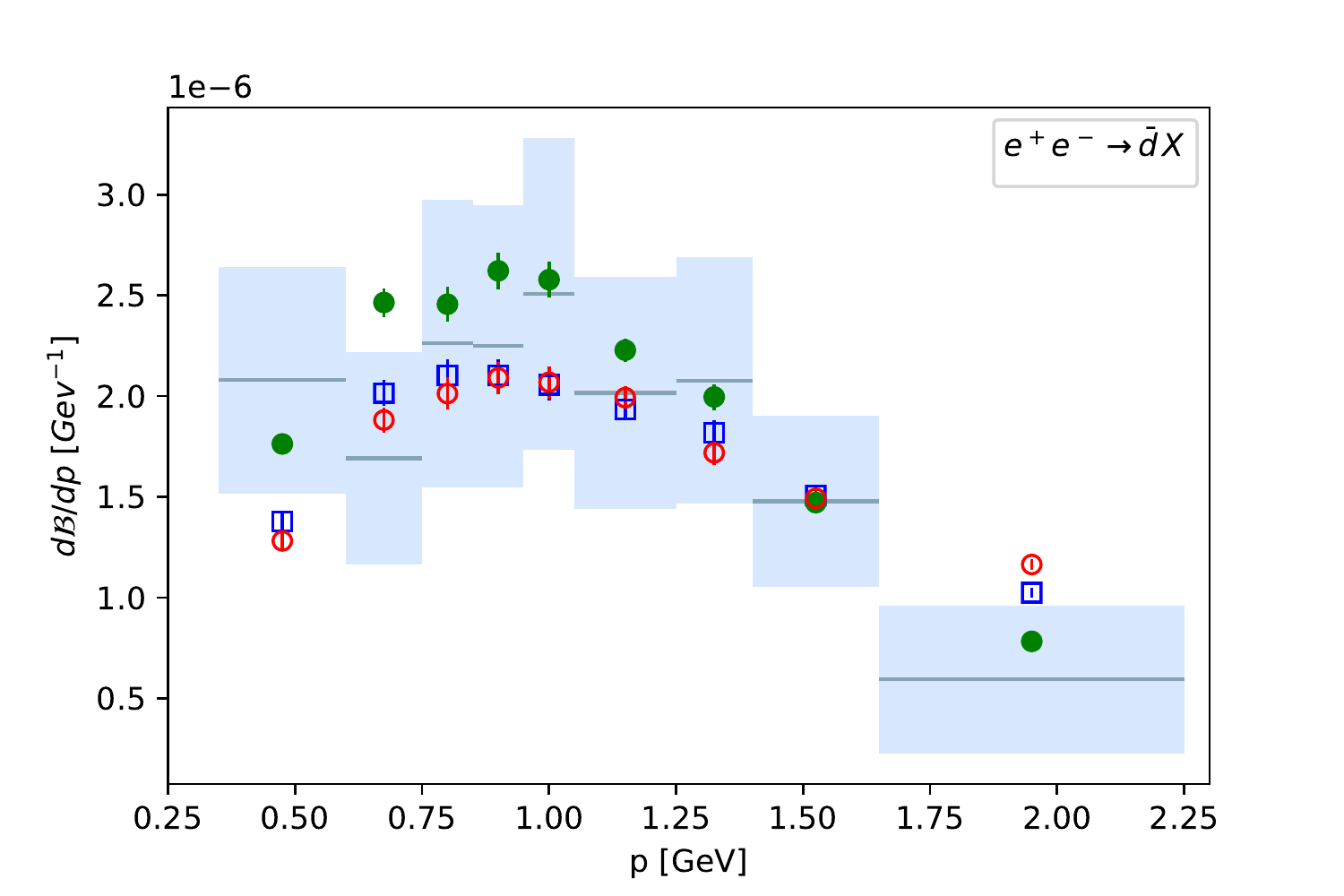}
    \caption{The momentum spectra of the $\bar{d}$ produced according to the simple coalescence model (blue squares), the cross section based model (green points) and the advanced coalescence model (red points), using the values reported in Table~\ref{tab:best_fit}. A comparison with the experimental results from BaBar (light blue boxes)~\cite{BaBar:2014ssg} and CLEO (black triangles)~\cite{CLEO:2006zjy} is done. Notice that error bars for experimental results are given by the quadratic sum of statistical and systematic uncertainties.}
    \label{fig:momentum_spectra}
\end{figure*}

Once the model parameters are fitted to the branching ratios, we compare the predicted $\bar{d}$ spectra with the observed one (Fig.~\ref{fig:momentum_spectra}). We found that all the three models are able to consistently reproduce the experimental spectra within their large  uncertainties.
Model A and B predict very similar spectra, which are on average slightly harder than the experimental ones, while model C generates spectra that are in average softer. This can be due to the contribution of $\bar N_1 \bar N_2 \to \pi\pi\bar{d}$, which on average produce softer $\bar{d}$. These features are present in all the processes we studied. 
More precise experimental data are needed to distinguish among different models.

\section{Study of the $d \bar{d}$ associated production}
\label{sec:ass_prod}
The $d \bar{d}$ associate production is potentially sensitive to effects not described by the two-nucleon coalescence. The CLEO collaboration reported the presence of a deuteron in 4 out of  338 events with a  $\bar{d}$ candidate~\cite{CLEO:2006zjy}. One $d$ candidate was identified as being produced by hadronic interactions on the detector material, while the other 3 are compatible with a prompt production in the interaction region. We estimate an average $70\%$ efficiency from the original publication, and  calculate the experimental ratio:
\begin{equation}\label{eq:associate_prod}
    \frac{\mathcal{B}(\Upsilon(1S)\rightarrow d\bar{d}X)}{\mathcal{B}(\Upsilon(1S)\rightarrow \bar{d}X)} = (13_{-7}^{+11}) \times 10^{-3}.
\end{equation}
To test this measurement we generate new MC samples consisting of $10^9$ events for each $\Upsilon(nS)$ and $3.69\times10^9$ events for the continuum, searching for events in them with a $d \bar{d}$ pair using all three models and fixing their parameters to the values in Table~\ref{tab:best_fit}. The resulting prediction on $\frac{\mathcal{B}(\Upsilon(1S)\rightarrow d\bar{d}X)}{\mathcal{B}(\Upsilon(1S)\rightarrow \bar{d}X)}$ is almost a factor $10$ below the value we calculate using the CLEO data, as reported in Table~\ref{tab:ass_predict}, which however corresponds to only $2$ standard deviations.
\begin{table}[th]
\begin{center}
\begin{tabular}{lccc}
	\hline
	 Process & Model A & Model B  & Model C\\
	\hline\hline
	$\frac{\mathcal{B}[\Upsilon(1S)\rightarrow d\bar{d} X]}{\mathcal{B}[\Upsilon(1S)\rightarrow \bar{d} X]}$ & $1.6\pm0.2$ & $1.3\pm0.2$  & $1.1\pm0.2$\\
	$\frac{\mathcal{B}[\Upsilon(2S)\rightarrow d\bar{d} X]}{\mathcal{B}[\Upsilon(2S)\rightarrow \bar{d} X]}$ & $1.0\pm0.2$ & $1.2\pm0.2$  & $1.1\pm0.2$\\
	$\frac{\mathcal{B}[\Upsilon(3S)\rightarrow d\bar{d} X]}{\mathcal{B}[\Upsilon(3S)\rightarrow \bar{d} X]}$ & $1.2\pm0.2$ & $1.0\pm0.2$  & $0.9\pm0.2$\\
	$\frac{\sigma(e^+e^-\rightarrow d\bar{d} X)}{\sigma(e^+e^- \rightarrow \bar{d} X)}$ & $0.3\pm0.2$ & $0.5\pm0.2$  & $0.4\pm0.2$\\
	\hline
\end{tabular}
\end{center}
\caption{Predictions for the ratios of $d\bar{d}$ production rate in $\Upsilon(nS)$ decays and $q\bar{q}$ fragmentation at $\sqrt{s} = 10.58$\gev in units of $10^{-3}$.}
\label{tab:ass_predict}
\end{table}
All models indicate that the production of $\bar{d} d$ pairs normalized to the single antideuteron is about 3 times larger in $\Upsilon(nS)$ decays than in the continuum.

\section{$H$-dibaryon production}
\label{sec:hdibaryon}
Several authors suggested the existence of an $H$-dibaryon: a six-quark, $udsuds$ state with quantum numbers  $I = 0$, $S=-2$ and $J^P = 0^+$. Depending on its mass, it could manifest as a deeply bound completely stable state, as a weakly decaying particle, or as a resonance decaying strongly.
For masses few\mev below twice the $\Lambda$ mass, the $H$ would behave as a $\Lambda\Lambda$ loosely-bound molecule and could be described with the same models used for the deuteron. 
This $H$ would predominantly decay via $\Delta S = +1$ weak transitions, for example to $\Lambda n$, $\Sigma^- p$, $\Sigma^0 n$ or $\Lambda p \pi^-$.
The observation of a very loosely bound $^{\phantom{\Lambda}6}_{\Lambda\Lambda}\text{He}$ hypernucleus has been used to set a lower limit to the two-$\Lambda$ binding energy $B_{\Lambda\Lambda}\lesssim 7.66\mev$~\cite{nagara}. More recently, the ALICE collaboration studied the $\Lambda\Lambda$ correlation in $pp$ and $p\text{Pb}$ collisions, further narrowing down limits to $B_{\Lambda\Lambda}=3.2^{+1.6}_{-2.4}{}^{+1.8}_{-1.0}\mev$~\cite{ALICE:2019eol}.
Lattice QCD extrapolations to physical pion masses by the NPLQCD collaboration reported a binding energy $B_H = 7.4 \pm 6.2\mev$~\cite{Beane:2011zpa}, and subsequent studies with a chiral constituent model constrained by $\Lambda N$, $\Sigma N$, $\Xi N$ and $\Lambda \Lambda$ cross sections found a $B_H$ value similar to the NPLQCD one~\cite{Carames:2012zz}. Other recent lattice QCD report a deeper $B_H = 19 \pm 10\mev$ at heavier pion masses~\cite{Francis:2018qch}.

Searches for a resonance decaying into $\Xi p \pi$ with mass around $2m_{\Lambda}$ have been performed by several experiments. The Belle collaboration, in particular, reported an upper limit for its production in $\Upsilon(1S, 2S)$ decays~\cite{Belle:2013sba}, but no theoretical estimates are available so far.
\begin{table}[b]
\begin{center}
\begin{tabular}{lcccc}
	\hline
	$B_{H}$ [\mevnospace] & $\Upsilon(1S)$ & $\Upsilon(2S)$ & $\Upsilon(3S)$ & $e^+e^- \to q \bar{q}$ \\
	\hline\hline
	0.25 & 1.7 & 1.4 & 1.2 & 0.18 \\
	0.50 & 2.4 & 1.9 & 1.8 & 0.26 \\
	1.00 & 3.4 & 2.8 & 2.6 & 0.37 \\
	2.00 & 4.7 & 3.7 & 3.6 & 0.51 \\
	3.00 & 5.7 & 4.5 & 4.3 & 0.60 \\
	4.00 & 6.5 & 5.4 & 5.0 & 0.71 \\
    \hline
\end{tabular}\end{center}
\caption{Predictions ($\times 10^{-7}$) for $H$ production rate in various processes using model A with $\sigma = 3.36$~fm and varying the dibaryon binding energy $B_{H}$.}
\label{tab:predictions_H}
\end{table}

We applied the results obtained from the study of the $\bar{d}$ production to calculate for the first time the production rate of this $H$-dibaryon in bottomonium decays and $e^+e^-$ collisions as a function of its binding energy.  We use only the simplest model (A) for this study, as model B and A turned out to give very similar results for the deuteron, and we lack of information about $\Lambda\Lambda \to H + X$ exclusive scattering channels to apply model C.
Inverting equation~\eqref{eq:sigma} we calculate the string fragmentation length corresponding to the fitted antideuteron coalescence momentum.
We use $k_\text{cut} = 73.4^{+1.3}_{-1.5}\mev$ from the simultaneous fit of the $\Upsilon(nS)$, and $B_d = 2.22\mev$, obtaining  $\sigma_s = 3.36^{+0.09}_{-0.10}$~fm.
We assume that this value can be used to describe also the formation of $\Lambda\Lambda$ pairs. To account for the uncertainty resulting from this choice we also present the result for $\sigma_s=2$~fm and $\sigma_s = 5$~fm, which are the physical limits of the parameter according to~\cite{Gustafson:1993mm}.
We calculate the $H$ coalescence momentum as a function of its binding energy $B_H$ as
\begin{equation}\label{eq:H_imp_rel}
    (2k^H_\text{cut})^3 = \frac{45}{\sqrt{\pi}} \sigma_s^{-2} \sqrt{m_{\Lambda} B_H},
\end{equation}
which has the same structure of~\eqref{eq:sigma} except for the numerical factor accounting for the different spin of the $d$ and $H$.
Equation~\eqref{eq:H_imp_rel} holds for loosely-bound, extended objects for which $B_H < 1/(m_{\Lambda} \sigma_s^2)$. For $\sigma_s=2$~fm the maximum $B_H$ allowed is 8.7\mev, for $\sigma_s= 5$~fm its maximum value drops to 1.4\mev.

\begin{figure}[t]
	\centering
    \includegraphics[width=\columnwidth]{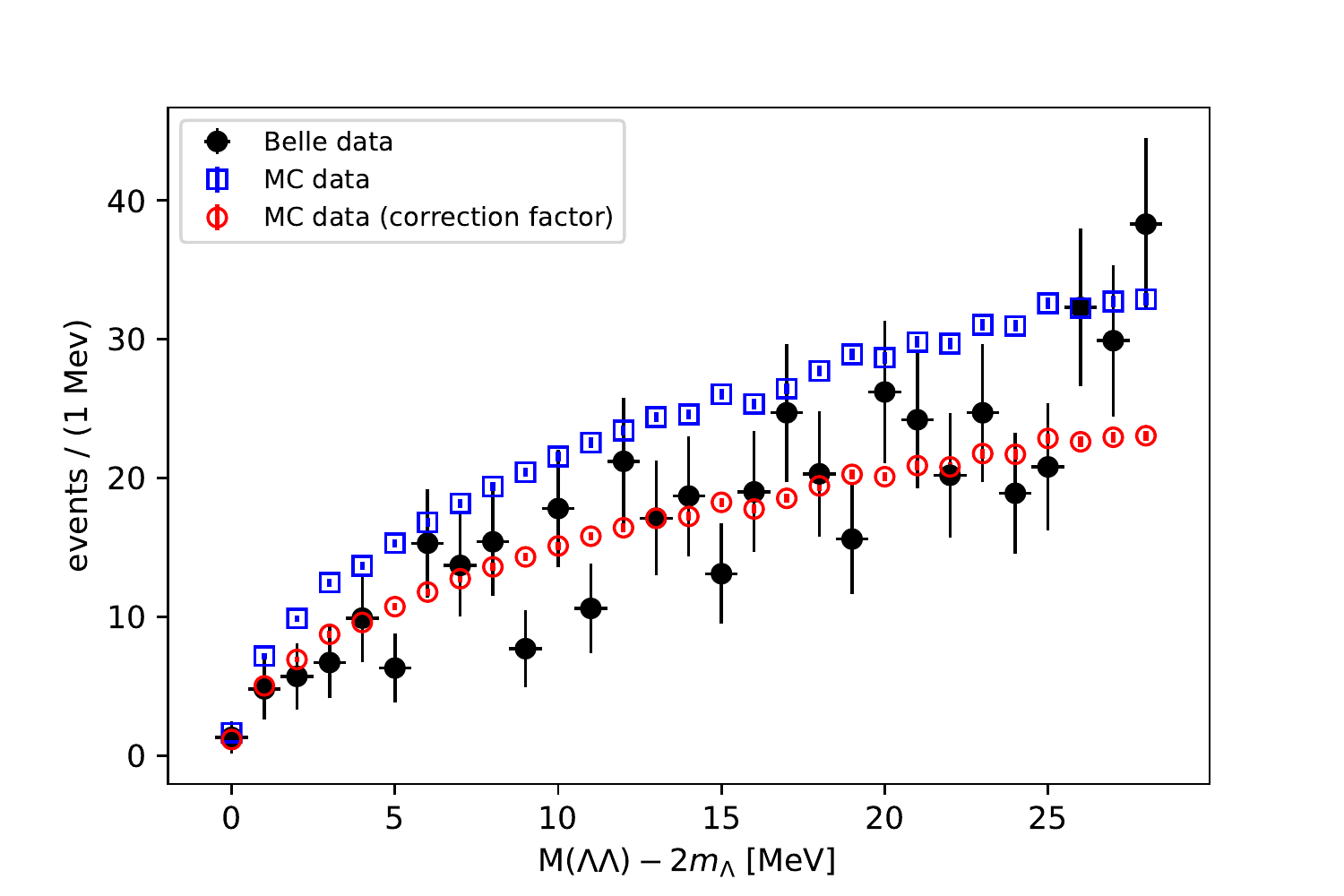}
    \caption{The $\Lambda\Lambda$ invariant mass distribution of reported in Ref.~\cite{Belle:2013sba} combining $\Upsilon(1S)$ and $\Upsilon(2S)$ experimental data (black points), compared with the distribution observed in the MC before (blue squares) and after (red circles) applying a scale factor of $0.71$.}
    \label{fig:ll_distr}
\end{figure}

We rely again on \pythia to model the two-baryon production and kinematics. We test our MC simulation against the  $\Lambda\Lambda$ invariant mass spectrum published by the Belle collaboration~\cite{Belle:2013sba} correcting for the experimental efficiency, the proportion between the $\Upsilon(1S)$ and $\Upsilon(2S)$ data samples, and the value of $\mathcal{B}(\Lambda \rightarrow p \pi^-)$. Our tuning overestimates the rate of $\Lambda\Lambda$ pairs at threshold by approximately $30\%$ but describes its shape correctly, as shown in Figure~\ref{fig:ll_distr}. Instead of re-tuning the MC we decide to apply a scale factor of $0.71$ to the simulation to match the experimental measurement.
In order to account for the feed-down from higher mass hyperons, in particular from the short-lived $\Sigma^0$, we drop the nonprompt $\Lambda$s produced from decays further than $\sigma = 3.36$~fm from the $e^+e^-$ interaction point.

The results are reported in Table~\ref{tab:predictions_H}. 
We found that the $H$ inclusive production rate is about two orders of magnitude lower than $\bar{d}$ one.  The $H$ is less likely to be produced from the continuum than in $\Upsilon(nS)$ by about one order of magnitude.

\begin{figure}[h]
	\centering
    \includegraphics[width=\linewidth]{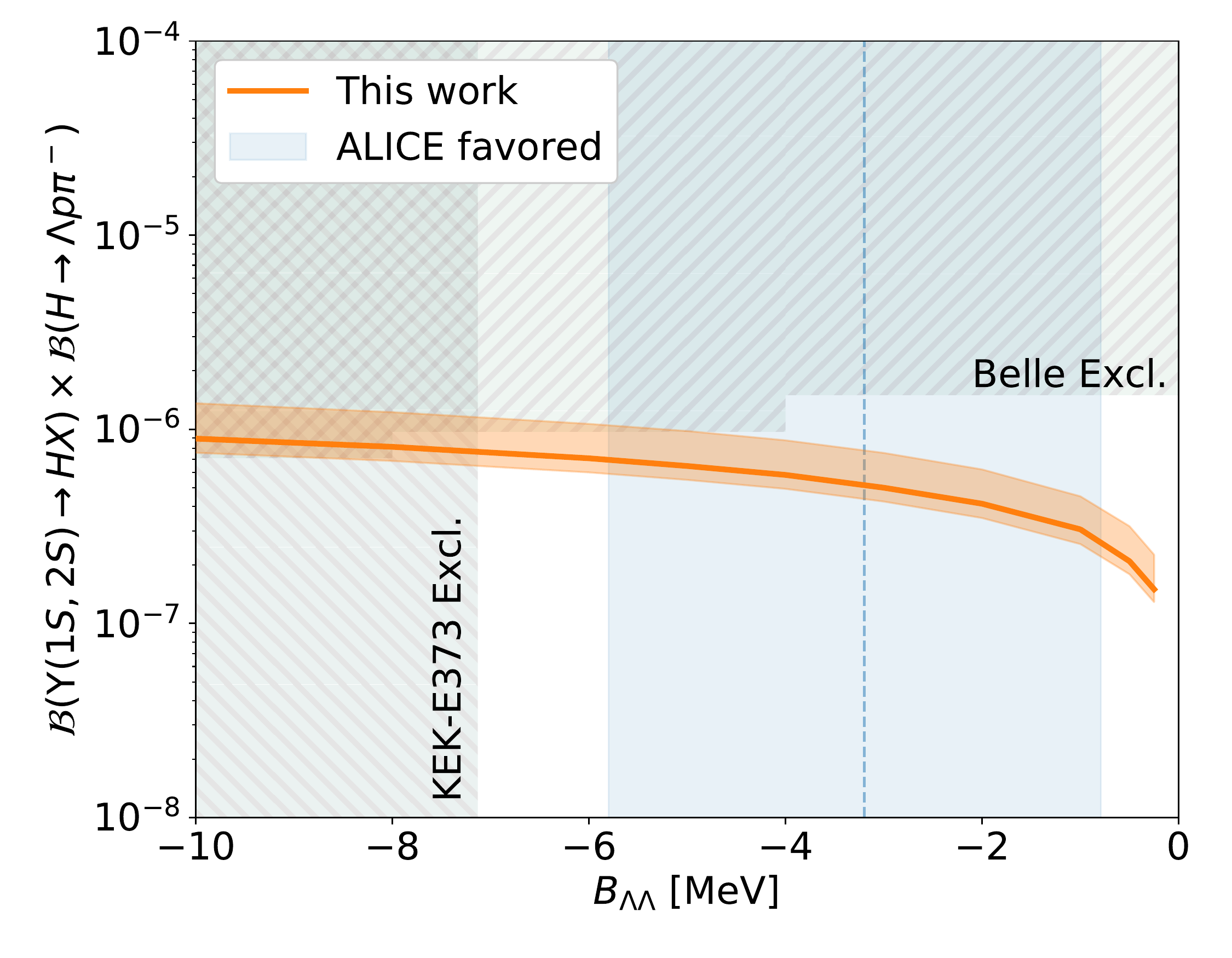}
    \caption{Comparison between our predictions for $\mathcal{B}(\Upsilon(nS) \rightarrow H X)\times  \mathcal{B}(H \rightarrow \Lambda p \pi^-)$ and the current experimental information.
    The error band on the prediction is obtained assuming a flat distribution for $\sigma_s$ within 2 and 5~fm. 
    The dashed areas are excluded by either direct searches or by the limit on the $\Lambda\Lambda$ binding energy. The most probable value for $B_{\Lambda\Lambda}$ obtained by ALICE is reported for reference~\cite{ALICE:2019eol}.}
    \label{fig:BR_sigma}
\end{figure}

Figure~\ref{fig:BR_sigma} shows a comparison between Belle upper limits and our prediction, obtained considering the $\Upsilon(1S)$ and $\Upsilon(2S)$ contributions proportionally scaled as Belle data.
The prediction is reliable above $B_H \gtrsim -3.1$~\mev, below which the loosely-bound hypothesis starts to fail.
In this region our predictions vary considerably as function of $\sigma_s$, remaining always below Belle limits.

\section{Hypertriton production}

Finally, we extend the approach of \hyperref[sec:hdibaryon]{previous section} to the hypertriton ($^3_{\Lambda}H$).
We model the $^3_{\Lambda}H$ as a deuteron-$\Lambda$ loosely-bound molecule, assuming a binding energy of $B_\text{ht}=0.41\mev$~\cite{STAR:2019wjm}.
Results on production rates are reported in Table~\ref{tab:hypertriton}, and show a five orders of magnitude suppression with respect to the antideuteron one, hinting to a much stronger suppression of hypertriton than the one observed in heavy ion collision, where the ratio is about $10^{-3}$~\cite{ALargeIonColliderExperiment:2021puh}.
\begin{table}[h!]
\begin{center}
\begin{tabular}{lc}
	\hline
	Process & Value \\
	\hline\hline
	 ${\cal B} [\Upsilon(1S) \rightarrow {_{\Lambda}^3H} + X]$ &  $2^{+2.6}_{-1.3}\times 10^{-10}$ \\
	 ${\cal B}[\Upsilon(2S) \rightarrow {_{\Lambda}^3H} + X]$ & 
	 $3^{+2.9}_{-1.6}\times 10^{-10}$  \\
	 ${\cal B}[\Upsilon(3S) \rightarrow {_{\Lambda}^3H} + X]$ & 
	 $2^{+3.6}_{-2.4}\times 10^{-10}$ \\
	 $\sigma[e^+e^- \rightarrow {_{\Lambda}^3H} + X]$ & 
	 $< 3\times10^{-4}$ fb\\
	\hline
\end{tabular}\end{center}
\caption{Predictions for $^3_{\Lambda}H$ production rate in the bottomonium region using model A. We fixed the $d$-$\Lambda$ binding energy to $B_\text{ht}=0.41\mev$~\cite{STAR:2019wjm}.}
\label{tab:hypertriton}
\end{table}

\section{Conclusions}

We presented a comprehensive test of the coalescence models for antideuteron production on the available samples of bottomonium decays and $e^+e^-$ collision in the $\sqrt{s} \approx 10.5\gev$ region. We performed a dedicated tuning of the \pythia generator to ensure the best possible description of the hadronization dynamics. We found that once the model parameters are fitted to reproduce the measured branching ratios, they all describe the observed antideuteron momentum spectrum within the experimental uncertainties. The current precision of the experimental measurements do not allow to falsify any of the tested model. A reduction of the total uncertainty by a factor three, which should be in reach of Belle~II experiment, would allow to distinguish between the coalescence and the cross section-based model.
We used the models to predict the $d\bar{d}$ inclusive production rate. Results are in mild disagreement with the measurement inferred from the CLEO data~\cite{CLEO:2006zjy}. A measurement of the $d\bar{d}$ production rate would be highly useful as a stringent test of the coalescence models.

We applied our findings to make the first estimate of production of a loosely-bound $H$-dibaryon at these energies. This reveals a suppression of one to two orders of magnitude compared to the correspondent $\bar{d}$ production rate. Our estimate for $\Upsilon(1S, 2S) \to H  X$ is lower than the upper limit set by the direct search at Belle~\cite{Belle:2013sba}, but likely within reach of Belle~II.

Finally, we presented the first prediction on hypertriton ($^3_{\Lambda}H$) production at these energies, modeling it as a deuteron-$\Lambda$ loosely-bound molecule. We found a five order of magnitude suppression with respect to the antideuteron production.

\begin{acknowledgments}
U.T. would like to thank Jonas Tjemsland and Alexander Kalweit for the interesting discussions during the MIAPP workshop {\it Anti-nuclei in the universe?}, and Roberto Mussa for bringing up the topic of antideuteron measurements at $B$-factories.
This work has been partially supported by the INFN  {\it ComPID} grant and by the Munich Institute for Astro- and Particle Physics (MIAPP) which is funded by the Deutsche Forschungsgemeinschaft (DFG, German Research Foundation) under Germany's Excellence Strategy –- EXC-2094 –- 390783311.
\end{acknowledgments}

%

\end{document}